\documentclass[11pt]{article}

\usepackage{graphicx}
\usepackage{multicol}
\usepackage{color}
\usepackage{epsfig,latexsym}
\usepackage{textcomp}

\definecolor{verdon}{cmyk}{1,0.5,1,0}
\definecolor{blue}{cmyk}{0.8,0.8,0,0.}
\definecolor{red}{cmyk}{0.2,1,1,0.0}

\def\lapprox{\mathrel{\mathop  {\hbox{\lower0.5ex\hbox{$\sim$}
\kern-1.1em\lower-0.7ex\hbox{$<$}}}}}
\def\gapprox{\mathrel{\mathop  {\hbox{\lower0.5ex\hbox{$\sim$}
\kern-1.1em\lower-0.7ex\hbox{$>$}}}}}

\begin{document}

\title{\color{verdon} 
A step toward CNO solar neutrinos detection in liquid scintillators}

\author{F.L. Villante$^{1,2}$, A. Ianni$^{2}$, F. Lombardi$^{1,2}$, G. Pagliaroli$^{2}$, F. Vissani$^{2}$
\vspace{0.5 cm}\\
{\small\em $^{1}$Universit\`a dell'Aquila, Dipartimento di Fisica, L'Aquila, Italy}\\
{\small\em $^{2}$INFN, Laboratori Nazionali del Gran Sasso, Assergi (AQ), Italy}}

\date{}

\maketitle

\def\abstractname{\color{red}\bf Abstract}
\begin{abstract}
{\footnotesize 
The detection of CNO solar neutrinos in ultrapure liquid scintillator detectors is limited by the background 
produced by Bismuth-210 nuclei that undergo $\beta$-decay to Polonium-210 with a lifetime of $\sim 7$ days. 
Polonium-210 nuclei are unstable and decay with a lifetime equal to $\sim 200$ days emitting $\alpha$ particles 
that can be also detected. 
In this letter, we show that the Bi-210 background can be determined 
by looking at the time evolution of $\alpha-$decay rate of Po-210,
provided that $\alpha$ particle detection efficiency is stable over the data acquisition period and external sources 
of Po-210 are negligible.
A sufficient accuracy can be obtained in a relatively short time. 
As an example, if the initial Po-210 event rate is $\sim 2000\,{\rm cpd/100 ton}$ or lower, 
a Borexino-like detector could start discerning  CNO neutrino signal from Bi-210 background in $\Delta t \sim 1$yr.
}
\end{abstract}

\newpage

\section{Introduction}

 One of the main goals of the present and next generation ultrapure
liquid scintillator detectors, such as KamLAND \cite{Kamland}, 
Borexino \cite{BXfirst}, SNO+ \cite{SNO+} and LENA \cite{LENA}, 
is the determination of the neutrino fluxes produced by the CNO cycle in the Sun.

Despite being sub-dominant in the Sun, the CNO cycle \cite{bethe} has a key role in 
astrophysics, being the prominent source of energy in more massive stars 
and in advanced evolutionary stages of solar like stars, see \cite{castellanietal}.
The evaluation of CNO efficiency is connected with various interesting problems,
like e.g. the determination of globular clusters age \cite{noiS14} from which we extract
a  lower limit to the age of the Universe.  At the moment, we still miss a direct observational 
evidence for CNO energy generation in the Sun. We only have a loose upper limit on CNO luminosity, $L_{\rm CNO}$, 
obtained by combining the results of the various solar neutrino experiments.
The most recent Solar Standard Model (SSM) \cite{Carlos} predicts the CNO contribution to the Sun's luminosity  to be equal to about 0.7\%. 
Experimental bounds report $L_{\rm CNO}<3$\% at 99\% C.L., see e.g. \cite{Gonzalez,StrumiaVis,PRLBorex}. 
The detection of CNO solar neutrinos would clearly provide a direct test of the CNO cycle efficiency.

The measurement of the CNO solar neutrino flux can also provide clues to
solve the so called ``solar composition problem''.
The flux  is, in fact, directly related to the abundance of 
carbon, nitrogen and oxygen in the Sun. The photospheric abundances 
of these (and other) heavy elements 
have been recently re-determined \cite{as05,as09,caffau09}, 
indicating that the sun metallicity is lower than previously
assumed \cite{gs98}. Solar models that incorporate these lower abundances 
are no more able to reproduce the helioseismic results.
Detailed studies have been done to solve this discrepancy, see e.g. \cite{Basu,noi},
but a definitive solution of this problem is still missing.

\begin{figure}[t]
\par
\begin{center}
\includegraphics[width=12.cm,angle=0]{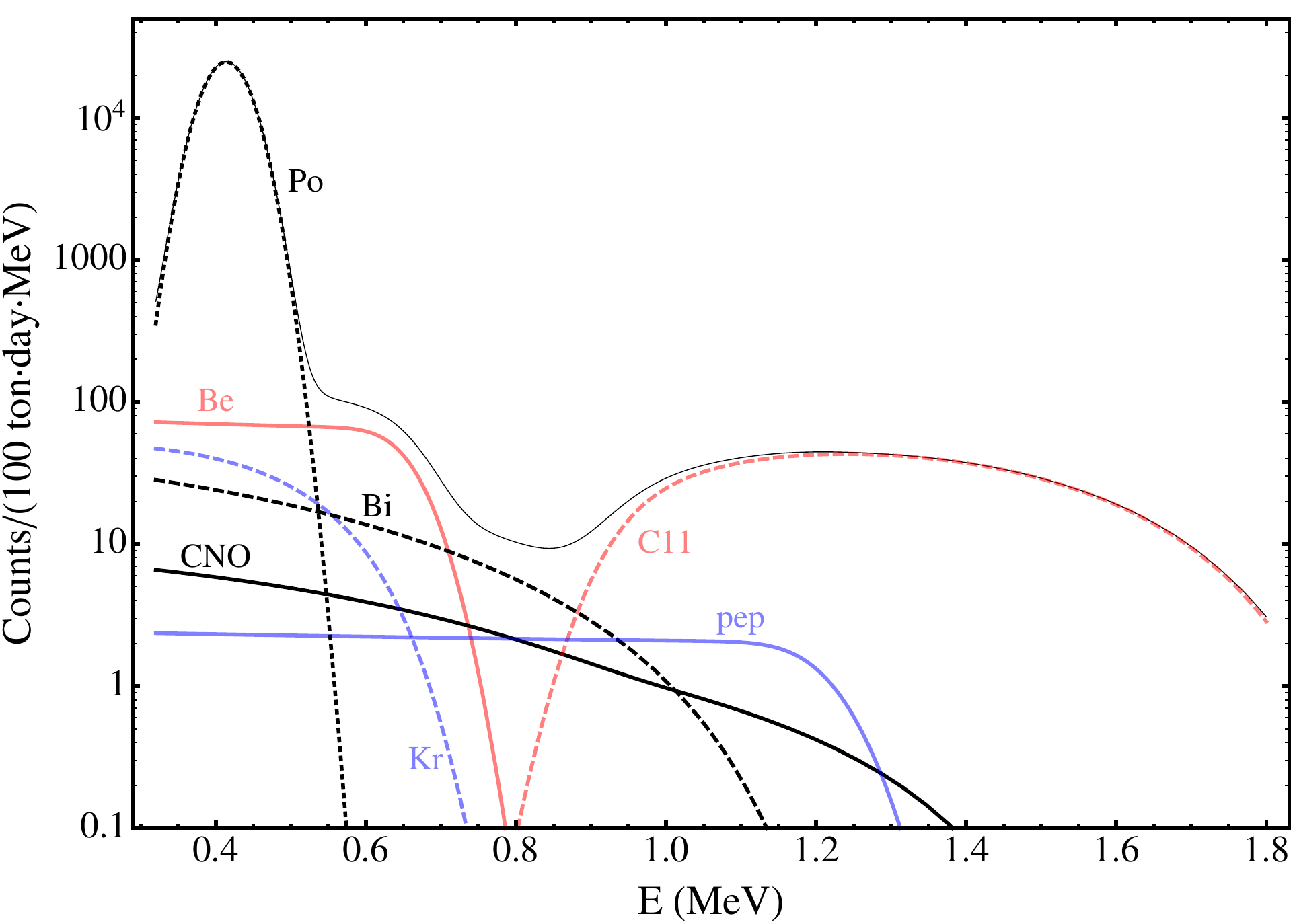}
\end{center}
\par
\vspace{-5mm} \caption{\em {\protect\small  The expected event spectrum in solar neutrino liquid scintillator detectors
calculated by assuming: the solar neutrino fluxes predicted by \cite{Carlos}; the oscillation parameters
corresponding to the LMA-MSW flavour oscillation solution \cite{PDG}; 
background levels and detector energy resolution comparable to those obtained by 
Borexino detector \cite{PRLBorex}. See sect.\ref{Implications} for details.}}
\label{Spectrum}
\end{figure}

The above points show the importance of CNO solar neutrinos detection which, 
however, is  a very difficult task.  Not only the flux is
relatively low, but also their energy is not large.
The neutrinos produced in the CNO cycle 
have continuous energy spectra with endpoints at about $\sim1.5$ MeV\footnote{
The dominant components of the CNO neutrino flux are the so called N-neutrinos 
produced by $^{13}{\rm N}\rightarrow ^{13}{\rm C}+e^+ + \nu_{\rm e}$ with an endpoint at $E_\nu = 1.20$ MeV
and the O-neutrinos produced by $^{15}{\rm O}\rightarrow ^{15}{\rm N}+e^+ + \nu_{\rm e}$ with an endpoint 
at $E_\nu = 1.73$ MeV.}.
Differently from the monochromatic Be and pep solar neutrinos, 
they do not produce specific spectral features that permit to 
extract them unambiguously from the background event spectrum
in high purity liquid scintillators. 
%
In particular, as it is shown in Fig.\ref{Spectrum}, the 
electrons produced by the
$\beta$-decay of Bismuth-210 to Polonium-210 
have a spectrum that is similar to that produced by 
CNO neutrinos. As a consequence, spectral fits are able to  determine only 
combined ``Bismuth+CNO'' contribution, as it is done e.g. by Borexino 
in \cite{BXfirst,PRLBorex}.

In order to remove this degeneracy, we propose a simple method
to determine the Bi-210 decay rate
which is based on the relationship between the Bi-210 and 
Po-210 abundances.  
Polonium-210, which is the Bismuth-210 daughter,
is unstable and decays with a lifetime $\tau_{\rm Po}\sim 200$ days emitting a
monochromatic $\alpha$
particle that can be easily detected. In the absence of Bismuth-210, 
the $\alpha-$decay rate of Po-210 nuclei 
follows the exponential decay law, $n_{\rm Po}(t) \propto \exp(-t/\tau_{\rm Po})$.
The deviations from this behaviour can be used to determine  
the $\beta-$decay rate of Bi-210 nuclei. The only required assumptions are that the
$\alpha$ particle detection efficiency is stable and external sources 
of Po-210 are negligible during the data acquisition period.

The plan of the paper is the following. In the next section, 
we determine the relationship between the 
$\alpha-$decay rate of Po-210, $n_{\rm Po}(t)$, 
and the $\beta-$decay rate of Bi-210, $n_{\rm Bi}(t)$.
In sect.\ref{Determination}, we estimate the accuracy $\Delta n_{\rm Bi}$ of the Bi-210 decay rate determination 
as a function of the detector mass $M$, the data acquisition period $\Delta t$ and the initial polonium decay rate $n_{\rm Po,0}$.
In sect.\ref{Implications}, we discuss the implications of the proposed approach for CNO neutrino signal extraction.
In sect.\ref{Conclusions}, we summarize our results.

\section{The relationship between Polonium and Bismuth}
\label{Relationship}
 
Bismuth-210 and Polonium-210 are both daughter of $^{238}{\rm U}$. 
Bismuth-210 is produced by the slow decay of Lead-210  which has a lifetime equal to $\tau_{\rm Pb}= 32.3 \; {\rm y}$.
It then undergoes a $\beta^-$ decay to Polonium-210:
\begin{equation}
^{210}{\rm Bi} \rightarrow  ^{210}\!{\rm Po} + e^{-} + \overline{\nu}_e
\end{equation}
 with a lifetime $\tau_{\rm Bi} = 7.232 \;{\rm d}$.
The electrons produced in the decay have a continuous spectrum 
with an endpoint $E_{\rm max}=1.16\, {\rm MeV}$ that lays over the event spectrum produced by CNO neutrinos, see Fig.\ref{Spectrum}.
In the analysis of the 192 days data \cite{PRLBorex}, Borexino obtained an event rate equal 
to about $20 \, {\rm cpd/100 tons}$ \cite{PRLBorex} for the combined contribution provided by
Bismuth+CNO neutrinos. For reference, we consider that the
expected CNO neutrino signal is equal to about 5~cpd/100ton as predicted by the SSM and the LMA-MSW scenario (see sect.\ref{Implications} for details). 

Polonium-210 is also unstable and undergoes $\alpha$ decay 
to Lead-206:
\begin{equation}
^{210}{\rm Po} \rightarrow  ^{206}\!{\rm Pb} + \alpha
\end{equation}
with a lifetime $\tau_{\rm Po} = 199.634\;{\rm  d}$.
The $\alpha$ particles produced in the decay have an energy $E_{\alpha}=5.3 \; {\rm MeV}$,
but they are observed at a much lower effective energy, due to the large 
quenching factor of alpha particles in liquid scintillators.
They produce a sharp peak superimposed to the low energy part of 
the $^{7}{\rm Be}$ neutrino event rate at a visible energy 
$E_{\rm vis}\simeq 0.5 \, {\rm MeV}$, see Fig.\ref{Spectrum}.

Naively, one could expect that secular equilibrium between 
Bismuth-210 and Polonium-210 is obtained, which would imply that the 
rate of $\beta$ decays of Bi-210 is equal to the rate of $\alpha$ decays of Po-210.
However, this is not the case in practical situations. 
Borexino, in fact, observes a Po-210 abundance which is about a 
factor 100 larger than what implied by secular equilibrium and that corresponds
 to an initial alpha decay rate of the order  $\sim 8\times10^{3} \, {\rm cpd/100 ton}$ \cite{TesteraPhysun2010}.
%
The large counting rate permits to determine the Polonium activity 
with a high accuracy. 

We remark that the $\alpha$ particles emitted in the decay, beside producing the very 
distinctive spectral feature shown in Fig.\ref{Spectrum}, can be directly identified by using a pulse shape discrimination. As it is shown in \cite{BXfirst,PRLBorex} a high purity liquid scintillator could have very good $\alpha-\beta$ discrimination properties. A figure-of-merit based on the Gatti discrimination technique \cite{Gatti} has been used in Borexino to produce an $\alpha$-subtracted spectrum where the Po-210 background is removed.

 The relationship between the Po-210 and the Bi-210 abundances can be
 quantified in a very simple way.  We have that:
\begin{equation}
\frac{d N_{\rm Po}}{dt} = - \frac{N_{\rm Po}(t)}{\tau_{\rm Po}} +
\frac{N_{\rm Bi}(t)}{\tau_{\rm Bi}} + S_{\rm Po}(t)
\label{start-eq}
\end{equation}
where $N_{\rm Po}(t)$ ($N_{\rm Bi}(t)$) is the number of Polonium-210 (Bismuth-210) 
nuclei per unit mass in the detector at a time $t$, 
and $S_{\rm Po}(t)$ indicates any possible external source of polonium-210 (i.e. not related to $^{238}{\rm U}$ decay chain).
The solution of Eq.~(\ref{start-eq}) can be written as:
\begin{equation}
N_{\rm Po}(t) = N_{\rm Po, 0} \, \exp(-t/\tau_{\rm Po}) + \tau_{\rm Po} \left\langle \frac{N_{\rm Bi}(t)}{\tau_{\rm Bi}} +S_{\rm Po}(t) \right\rangle
\end{equation} 
where $N_{\rm Po, 0}$ is the number of Po-210 nuclei per unit mass at the time $t=0$ and 
the symbol $\langle f(t) \rangle$ indicates the time ``average'':
\begin{equation}
\langle f(t) \rangle = \frac{1}{\tau_{\rm Po}} \int_{0}^{t} dt' \;f(t-t') \exp(-t'/\tau_{\rm Po})
\end{equation} 
for the generic function $f(t)$.

We assume that, during a data acquisition period, after liquid scintillator purification, 
external sources of Polonium can be neglected, i.e. $S_{\rm Po}(t)\simeq 0$.
In this assumption, 
the $\alpha-$decay rate of
Polonium and the $\beta-$decay rate of Bismuth, given by:
\begin{eqnarray}
\nonumber
n_{\rm Po}(t) &\equiv&  N_{\rm Po}(t)/\tau_{\rm Po}\\
n_{\rm Bi}(t) &\equiv& N_{\rm Bi}(t)/\tau_{\rm Bi}
\end{eqnarray}
respectively,
follow the relation:
\begin{equation}
n_{\rm Po}(t)= n_{\rm Po, 0}  \, \exp(-t/\tau_{\rm Po}) + \langle n_{\rm Bi} (t) \rangle  \, .
\label{eq-general}
\end{equation}
The above equation provides all the ingredients that are necessary for our analysis.
It shows that, if we are able to measure $n_{\rm Po}(t)$ as a function of time 
with a high accuracy, we can provide an estimate for $n_{\rm Bi}(t)$, which is the main background 
for CNO neutrinos detection.

\section{Determination of the Bi-210 event rate}
\label{Determination}

The main observation of this letter is that, due to the high statistics of Po-210 events, 
the beta-activity of Bi-210 nuclei can be extracted from the time evolution of Polonium-210
counting rate in a relatively short time (i.e.~comparable with Polonium lifetime).
It is not necessary, in particular, to wait that secular equilibrium between Polonium-210 and Bismuth-210 
is reached, which would require much longer times.

\begin{figure}[t]
\par
\begin{center}
\includegraphics[width=8.cm,angle=0]{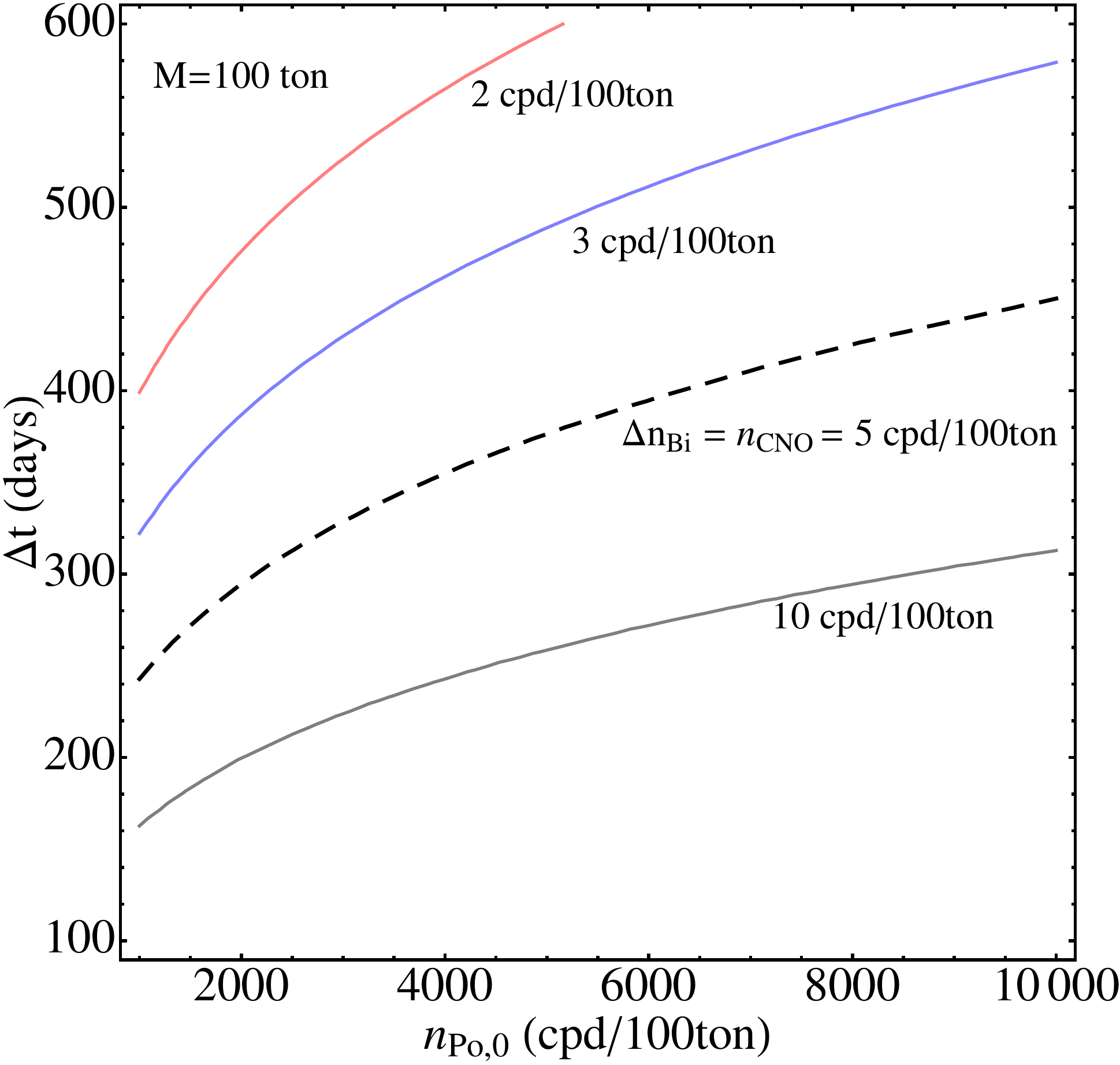}
\includegraphics[width=8.cm,angle=0]{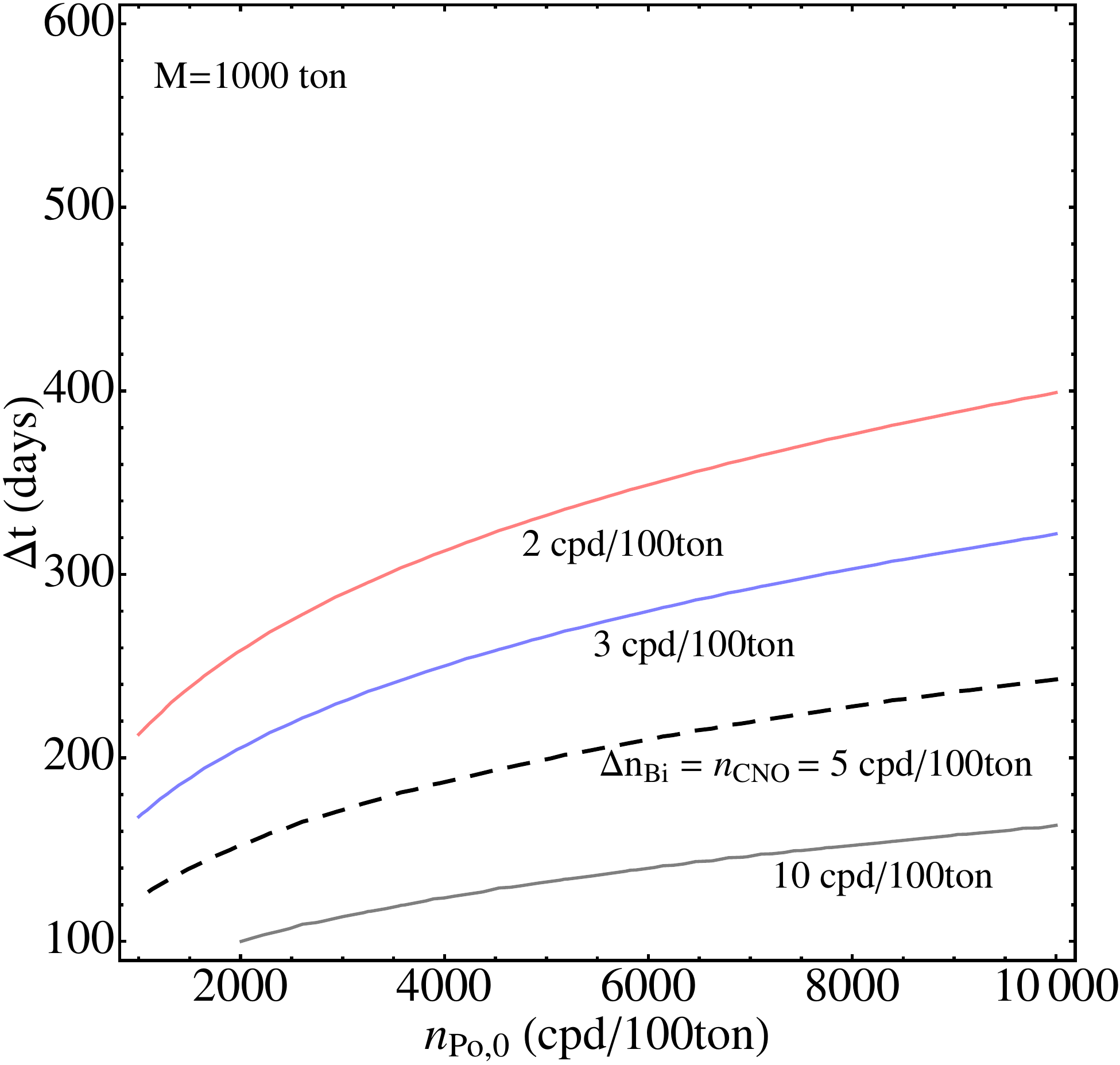}
\end{center}
\par
\vspace{-5mm} \caption{\em {\protect\small  The $1\sigma$ expected accuracy $\Delta n_{\rm Bi}$ in Bi-210 event rate determination.}}
\label{ExpectedAccuracy}
\end{figure}

This point can be quantified with straightforward analytical
considerations by making the following exercise. 
We assume that the Bi-210 event rate is constant\footnote{This hypothesis corresponds to assuming that
there are no sources of Lead-210 and Bismuth-210 in the detector and it is done here
only to consider a specific situation and to illustrate quantitatively the proposed method. 
We remark that, in the analysis  
of the experimental data, it is not necessary to make {\em a-priori} 
assumptions on the time evolution of $n_{\rm Bi}(t)$ since this can be determined 
from the experimental data themselves, as it is discussed in sect.\ref{Implications} and appendix.}, 
obtaining from Eq.(\ref{eq-general}):
\begin{equation}
n_{\rm Po}(t)= \left[ n_{\rm Po, 0} - n_{\rm Bi} \right] \, \exp(-t/\tau_{\rm Po}) + n_{\rm Bi} \,.
\label{eq-final}
\end{equation} 
We imagine to collect the Po-210 events\footnote{The Po-210 counting rate can be determined 
either by fitting the low energy part of the event spectrum or by selecting Po-210 events
by pulse shape discrimination tecnique, see \cite{Gatti}. The optimal approach has to be choosen on 
experimental basis.} 
for a total period  equal to $\Delta t $. We then divide 
this time interval into two large bins from $[0,\Delta t/2]$ and 
$[\Delta t /2, \Delta t]$.  According to the time evolution implied 
by Eq.(\ref{eq-final}), the number of events collected in the two bins are:
\begin{eqnarray}
N_{1} &=& \varepsilon \, M\,\left[ \left (n_{\rm Po, 0} - n_{\rm Bi} \right) \tau_{\rm Po} 
\left( 1 - e^{-\frac{\Delta t}{2 \tau_{\rm Po}}}\right) + n_{\rm Bi} \, \frac{\Delta t}{2} \right]
\label{bin1}
\\
N_{2} &=& \varepsilon \, M\,\left[ \left (n_{\rm Po, 0} - n_{\rm Bi} \right) \tau_{\rm Po}\, e^{-\frac{\Delta t}{2 \tau_{\rm Po}}}
\left( 1 - e^{-\frac{\Delta t}{2 \tau_{\rm Po}}}\right) + n_{\rm Bi} \, \frac{\Delta t}{2} \right]
\label{bin2}
\end{eqnarray}
where $M$ indicates the detector mass and $\varepsilon$ indicates the detection efficiency (averaged over the integration period)  
that we assume not to vary significantly.

In the absence of the Bismuth-210 contribution, the ratio of events collected in the two
bins is equal to $N_{2}/N_{1} = \exp\left(-\Delta t / 2 \tau_{\rm Po} \right)$. The deviations of $N_{2}/N_{1}$ from this value can be used to  
measure the beta activity of Bi-210. We have, in fact:
\begin{equation}
\left[N_{2}-N_{1}  \, e^{-\frac{\Delta t}{2 \tau_{\rm Po}}}  \right] = \varepsilon \, M \, n_{\rm Bi} \; \frac{\Delta t}{2} \; \left( 1 - e^{-\frac{\Delta t}{2 \tau_{\rm Po}}}\right) \,.
\end{equation}
By propagating the statistical errors $\Delta N_{2} = \sqrt{N_{2}}$ and $\Delta N_{1}= \sqrt{N_{1}}$, one is able 
to estimate the accuracy $\Delta n_{\rm Bi}$ of the determination of the Bi-210 decay rate. In the 
assumption that $n_{\rm Po, 0} \gg n_{\rm Bi}$, we obtain:
\begin{equation}
\Delta n_{\rm Bi} \simeq \sqrt{\frac{n_{\rm Po ,0}}{ \tau_{\rm Po} \, M}} \; f(\Delta t)
\label{Accuracy}
\end{equation}
where we considered $\varepsilon\sim 1$. The function $f(\Delta t)$ is explicitly
given as:
\begin{equation}
f(\Delta t) = \left(\frac{2\, \tau_{\rm Po}} {\Delta t}\right)\; e^{-\frac{\Delta t}{4 \tau_{\rm Po}}}\;\sqrt{\frac{1+e^{-\frac{\Delta t}{2 \tau_{\rm Po}}}}{1- e^{-\frac{\Delta t}{2 \tau_{\rm Po}}}}}\, .
\end{equation}
We see that the relevant parameters are the initial Po-210 event rate $n_{\rm Po, 0}$, the detector mass $M$ 
and the total time of the measure $\Delta t$. In Fig.\ref{ExpectedAccuracy}, we show the iso-countour lines for $\Delta n_{\rm Bi}$
in the plane $(n_{\rm Po,0},\Delta t)$ for two values of the detector mass  $M=100 \; {\rm ton}$ and 
$M= 1000 \;{\rm ton}$. For an initial Polonium activity equal to $n_{\rm Po,0}= 2000 \; {\rm cpd/100 \; ton}$, we reach 
an accuracy comparable to the expected CNO signal, $n_{\rm CNO} \simeq 5 \; {\rm cpd/100 \; ton}$, after a time 
$\Delta t \sim 300 d$ ($\Delta t \sim 150 d$) for a 100 ton (1000 ton) detector.
We remark that an important assumption in the above analysis is that the detection efficiency $\varepsilon$ is known and
stable over the period of the measure. Variations and/or errors in $\varepsilon$ clearly provide additional sources of 
uncertainties.

 These analytic estimates have been checked numerically 
by simulating random Po-210 events, according to the time evolution expressed by Eq.(\ref{eq-final}), 
in detectors with masses $M=100\,{\rm ton}$ and $M=1000\,{\rm ton}$. 
We assumed that the Bi-210 event rate is equal to $n_{\rm Bi} = 20\; {\rm cpd/ 100 ton}$, 
while the initial Po-210 event rate have been varied in the range $n_{\rm Po, 0} = 2000-8000 \; {\rm cpd/ 100 ton}$ 
in order to study the dependence of our final results on the initial polonium contamination.  
We binned the data over five days and then extracted $n_{\rm Bi}$ and $n_{\rm Po,0}$ by 
fitting the data over an observation period $\Delta t$ with the functional form in Eq.(\ref{eq-final}).
 The accuracy $\Delta n_{\rm Bi}$ for the reconstructed  Bi-210 event rate is reported in tab.\ref{tab1}. 
 We see that the numerical results agree within $\sim 15\%$ 
with the analytical expression (\ref{Accuracy}).

\begin{table}[t]
\begin{center}{
\begin{tabular}{l |ccc|ccc}
                 &           &  $M= 100 \, {\rm ton}$ &        &    &  $M= 1000 \, {\rm ton}$ & \\
\hline
                 & 0.5 year &  1 year & 1.5 year & 0.5 year &  1 year & 1.5 year \\

\hline
$n_{\rm Po, 0}   =2000 $ & 10 & 2.9 & 1.3 & 3.3 & 0.9  & 0.4\\ 
$n_{\rm Po, 0}   =4000 $ &  15 &  4.1 & 1.8 & 4.6 &  1.3  & 0.6\\
$n_{\rm Po, 0}   =8000 $  &  $> 20$ &  5.7 & 2.5 & 5.6 & 1.8 & 0.8\\
\hline
\end{tabular}
}\end{center}\vspace{0.4cm} \caption{\em {\protect\small The  absolute uncertainty $\Delta n_{\rm Bi}$ in the Bi-210 event rate determination, 
reconstructed by performing a fit over an observation period $\Delta t$ to simulated data. All rates are expressed in ${\rm cpd/ 100 ton}$. 
We assume that the true Bi-210 event rate is $n_{\rm Bi}= 20\, {\rm cpd/100 ton}$. 
\label{tab1} 
}}\vspace{0.4cm}
\end{table}

\section{Implications for CNO solar neutrino extraction}
\label{Implications}

 The possibility to determine the Bi-210 event rate represents an important step forward
in the direction of CNO solar neutrino detection, as can be immediately 
understood by looking at the 
expected event spectrum in liquid scintillator detectors in Fig.\ref{Spectrum}.
The spectrum was calculated by assuming the solar neutrino fluxes
predicted by \cite{Carlos} and the oscillation parameters 
corresponding to the LMA-MSW solution of the solar neutrino problem given by \cite{PDG}.
The integrated rates are $n_{\rm Be} = 50\, {\rm cpd/100 ton}$ from beryllium neutrinos, 
$n_{\rm pep}= 2.7\, {\rm cpd/100 ton}$ for pep neutrinos and $n_{\rm B}= 0.48\, {\rm cpd/100 ton} $ 
for boron neutrinos. The CNO neutrino signal is taken as 
$n_{\rm CNO}=  5.1\, {\rm cpd/100 ton}$ that corresponds to the prediction 
obtained by using the high photospheric metal abundances of \cite{gs98}. 
For comparison, by using the low photospheric values of \cite{as09} one obtains
the lower rate $n_{\rm CNO}= 3.6 \; {\rm cpd/100 ton}$.
The  background levels have been estimated considering the Borexino results as a reference (see e.g. \cite{PRLBorex}).
The Kripton-85 and Carbon-11 event rate are chosen as $n_{\rm Kr}= 25 \, {\rm cpd/100 ton}$ 
and $n_{\rm C11}= 25 \, {\rm cpd/100 ton}$, respectively.
The Bi-210 event rate is taken as $n_{\rm Bi} = 20 \, {\rm cpd/100 ton}$, while the (initial) Po-210
event rate is chosen as $n_{\rm Po, 0} =  2000 \, {\rm cpd/100 ton}$.
This value is lower than what initially obtained in Borexino \cite{TesteraPhysun2010} but we believe 
that it represents a reasonable goal for next runs and next generation experiments. 
Finally, we modelled the detector energy resolution by a Gaussian function with an energy dependent width  
$\sigma(E)=0.05 \, {\rm MeV} \, \sqrt{E/ 1 \; {\rm MeV}}$ \cite{PRLBorex}.

\begin{figure}[t]
\par
\begin{center}
\includegraphics[width=8.cm,angle=0]{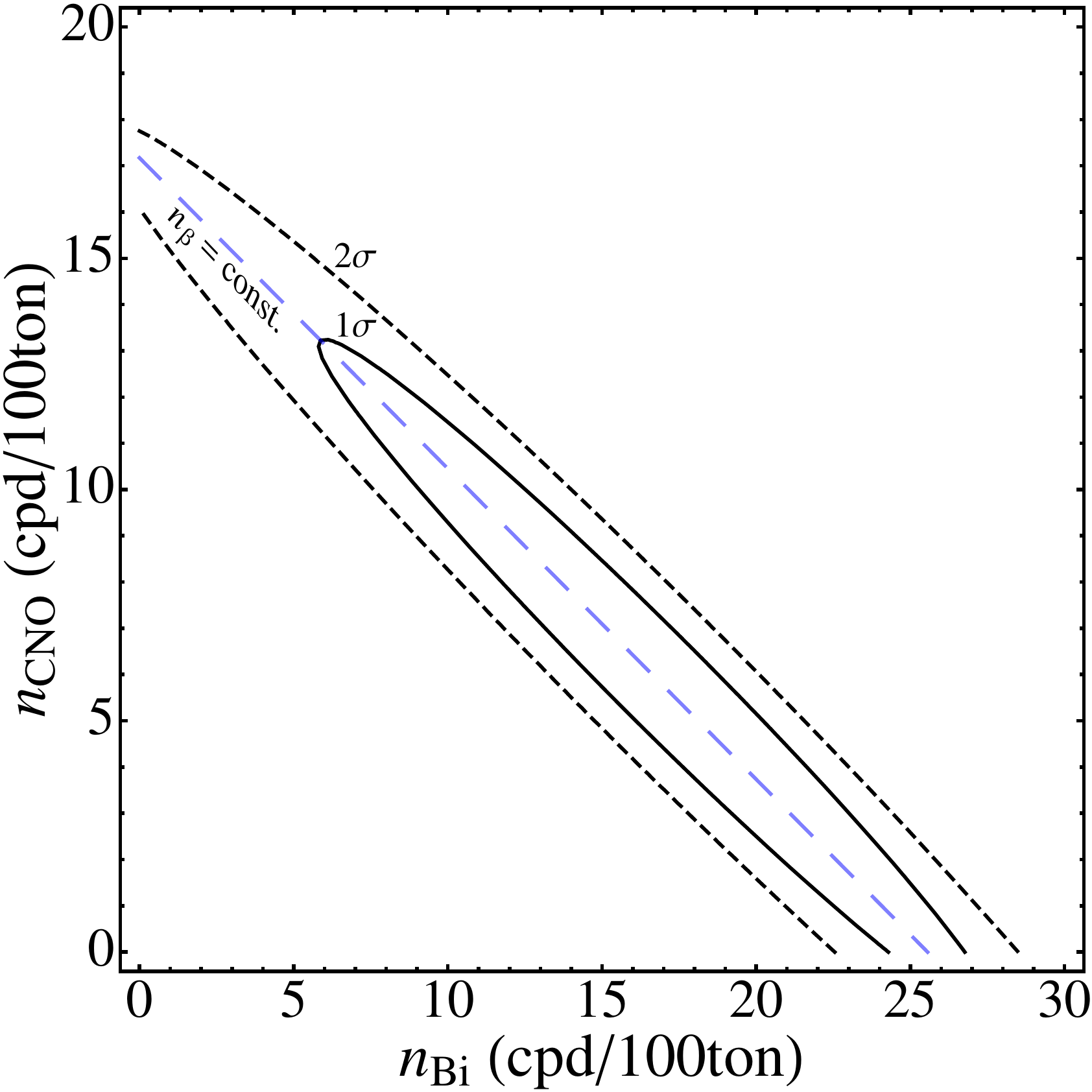}
\includegraphics[width=8.cm,angle=0]{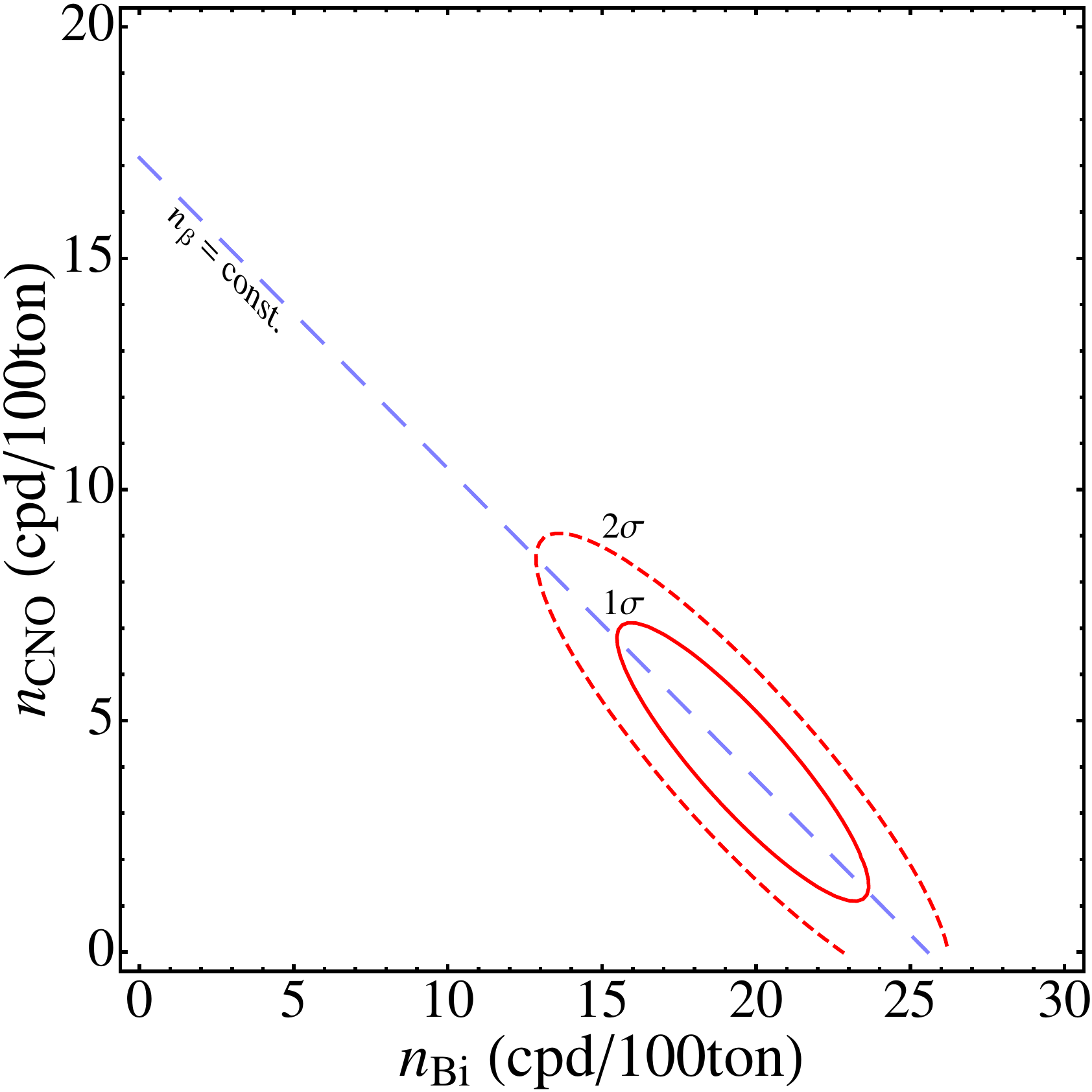}
\end{center}
\par
\vspace{-5mm} \caption{\em {\protect\small  The 1-$\sigma$ and 2-$\sigma$ allowed regions obtained by a fit to 
simulated data, assuming a detector mass $M=100 \, {\rm ton}$ and an observation period $\Delta t =1 \, {\rm yr}$. 
In the left panel, we consider only information contained in the energy distribution of the events. 
In the right panel, we perform a fit in the domain
of energy and time, taking into account the relationship between Po-210 and Bi-210 expressed by Eq.(\ref{eq-final}).
The blue dashed line corresponds to the condition $n_\beta = {\rm const}$, see Eq.(\ref{nbeta}).}}
\label{CNOestimate}
\end{figure}

We see that the CNO neutrino fluxes are not expected to produce 
significant and/or recognizable features anywhere in spectrum 
(see Fig.\ref{Spectrum}).
As a consequence, spectral fits are unable to constrain the CNO signal. 
This is emphasized in the left panel of Fig.\ref{CNOestimate}, where we 
display the results of a spectral fit to simulated data in the plane $(n_{\rm Bi}, n_{\rm CNO})$. 
In the simulation, we considered a detector mass $M=100\;{\rm ton}$ and 
a total observation period equal to $\Delta t= 1\, {\rm yr}$,
during which all the components have been assumed to be constant in time 
except the Po-210 contribution that was modulated in time according to Eq.(\ref{eq-final}). 
In the fit, we considered all the signals and backgrounds as free parameters,
except the pp, pep and Boron neutrinos contributions which have been fixed to their expected values.
We see from Fig.\ref{CNOestimate} that  the CNO signal is basically unconstrained.
 The origin of the strong anti-correlation 
with the Bi-210 background
is understood on a quantitative basis by considering that the Bi-210 and CNO components
give the dominant contribution to the spectrum only in a narrow energy range
at $E\sim 0.8 \, {\rm MeV}$. The signal produced in this window  is proportional to the combination:
\begin{equation}
n_{\beta} = n_{\rm CNO} + \xi\; n_{\rm Bi}
\label{nbeta}
\end{equation}
where the factor $\xi=0.67$ is the ratio between the normalized Bi-210 and CNO spectra evaluated at $E\sim 0.8$ MeV.
The data allow, thus, a good determination of $n_{\beta}$ with a poor reconstruction 
of  $n_{\rm CNO}$ and $n_{\rm Bi}$.

The degeneracy between Bi-210 and CNO neutrinos is removed when we use the time evolution
of the Po-210 contribution to determine the Bi-210 background.  
By performing a fit to the simulated data {\em in the domain of energy and time}
and taking advantage of Eq.(\ref{eq-final}) between Po-210 and Bi-210 rates, we obtain the considerable improvement in the CNO signal determination 
that is shown in the right panel of Fig.\ref{CNOestimate}\footnote{This plot has been obtained 
by fitting simultaneously the time and energy distribution of simulated data without considering
the possibility to identify Po-210 events by pulse shape discrimination \cite{Gatti}, since
this additional information do not introduce relevant improvements in our analysis. 
This possibility, however, should not be overlooked because it may became important in the analysis 
of real data, depending on the experimental conditions.}.
The final uncertainty is $\Delta n_{\rm CNO}\simeq 2.1 \,{\rm cpd/100 ton}$ and it is essentially
determined by the accuracy of the Bi-210 event rate determination according to 
$\Delta n_{\rm CNO} \sim \xi \; \Delta n_{\rm Bi} $, as can be estimated 
by using the $\Delta n_{\rm Bi}$ values given in tab.\ref{tab1} and/or in Eq.(\ref{Accuracy}).

The above results indicate that present generation liquid scintillator experiment, 
like e.g. Borexino, already have 
the potential to probe the CNO neutrino flux, provided that they are stable for sufficiently
long time ($\sim 1 {\rm yr}$) and/or the initial polonium contamination can be made sufficiently low.
Considering, moreover, that the uncertainty 
scales as $\Delta n_{\rm Bi} \propto \sqrt{n_{\rm Po,o}/M}$, we note that: 
{\em i)} future Kton-scale detectors, like e.g. SNO+, will be able to start discriminating 
between high and low metallicity solar models;
{\em ii)} the initial Po-210 contamination $n_{\rm Po,0}\sim 8000 \, {\rm cpd/100 ton}$ reported Borexino by \cite{TesteraPhysun2010}
corresponds, anyhow, to $1\sigma$ extraction of CNO solar neutrinos.

A final remark is important. 
In our analysis, we assumed that the Bi-210 event rate 
is constant in time. This assumption was done to consider a specific situation but
it is not necessary in the analysis of real data, since the time evolution of $n_{\rm Bi}(t)$
 can be determined from the experimental data themselves.
A simple strategy can be the following. By performing spectral fits to the signal at 
different times $t$, we determine the quantity $n_{\beta}(t)$ defined in Eq.(\ref{nbeta}).
This allow us to obtain Bi-210 rate as a function of time, according to:
\begin{equation}
n_{\rm Bi}(t) = \frac{1}{\xi}\left[n_{\beta}(t) - n_{\rm CNO}\right]
\label{nbi}
\end{equation}
where the unknown factor $n_{\rm CNO}$ is constant in time 
(we neglect the small seasonal modulation of neutrino fluxes due to earth orbit eccentricity).
The presence of a time varying 
$n_{\rm Bi}(t)$ does not invalidate the possibility to perform our analysis, 
since the only requisite for the validity of Eq.(\ref{eq-general}) is that 
there are no external Polonium-210 sources. We can use, in fact, Eq.(\ref{nbi}) in Eq.(\ref{eq-general})
to obtain the following expression:
\begin{equation}
n_{\rm Po}(t) - \langle n_{\beta} (t) \rangle  = \left[n_{\rm Po, 0} + n_{\rm CNO} \right] \, \exp(-t/\tau_{\rm Po})  - n_{\rm CNO}
\label{eq-td}
\end{equation}
that allows to determine $n_{\rm CNO}$ from a two parameters fit to the ``observable'' quantity 
$n_{\rm Po}(t) - \langle n_{\beta} (t) \rangle$ with no additional assumptions. 
Alternatively,  we can compare the behaviour of $n_{\beta}(t)$ with expectations. A constant value is 
expected when external sources of Bismuth-210 and Lead-210 are negligible. 
A linear growth of $n_{\rm Bi}(t)$ (and thus of $n_\beta(t)$) is, instead, produced by a slow increase of the Radon
contamination of the detector (see appendix for details). We can
assume an arbitrary functional form for $n_{\rm Bi}(t)$ that agrees with the observed $n_{\beta}(t)$ 
and fit the data in the time and energy domain, 
using Eq.(\ref{eq-general}) in place of Eq.(\ref{eq-final}) to discriminate Bi-210
from CNO neutrino contribution.

\section{Conclusions}
\label{Conclusions}

In this letter we have proposed a simple and general method that allows to measure the flux of CNO neutrinos in massive high radio-purity liquid scintillator detectors by means of 
determining the Bi-210 background.
We summarize here the main points of our analysis:\\
{\em i)} We have discussed the relationship between Polonium-210 and Bismuth-210, see Eq.(\ref{eq-general}), showing that 
the deviations of the $\alpha-$decay rate of Po-210 from
the normal exponential decay law, $n_{\rm Po}(t)\propto \exp(-t/\tau_{\rm Po})$,
can be used to determine  the beta activity of Bi-210.\\
{\em ii)} We have estimated the expected accuracy of the Bi-210 event rate determination 
as a function of the relevant parameters that are the initial polonium contamination of the detector, 
the detector mass and the total observational time. Our results are summarized by Eq.(\ref{Accuracy}) and/or tab.\ref{tab1}. 
A good accuracy can be obtained in relatively short time, comparable to the Po-210 lifetime $\tau_{\rm Po}\sim 200\, {\rm d}$.\\ 
{\em iii)} We have discussed the implications of our approach for CNO solar neutrino detection. We have shown that, 
if the initial Po-210 event rate is $\sim 2000\,{\rm cpd/100 ton}$ or lower, a Borexino-like detector could start 
discerning  CNO neutrino signal in $\Delta t \sim 1$yr. Future Kton-scale detectors, like e.g. SNO+, in the same time interval, 
could begin to discriminate between high and low metallicity solar models.

The determination of the Bi-210 event rate could represent an important step toward
CNO solar neutrinos detection in high purity liquid scintillator experiments. 
The only required assumptions are that $\alpha-$particle detection efficiency 
is stable over the data acquisitions period and external sources of Po-210 are negligible. A special 
effort on the experimental side in this direction is worthwhile in the next future.

\newpage


\appendix\section{Lead-Bismuth-Polonium 210 decay chain}

The decay chain  is described by the system of 
differential equations:
\begin{equation}
\begin{array}{l}
\frac{d N_{1}(t)}{dt} =   -\frac{N_1(t)}{\tau_1} +S_1(t) \\
\frac{d N_{2}(t)}{dt} =   -\frac{N_2(t)}{\tau_2} + \frac{N_1(t)}{\tau_1} +S_2(t) \\
\frac{d N_{3}(t)}{dt} =   -\frac{N_3(t)}{\tau_3} + \frac{N_2(t)}{\tau_2} +S_3(t) 
\end{array}
\end{equation}
where, to keep notation simple, we use the index 1 is for Lead-210, the index 2 for Bismuth-210 and the index 3 for Polonium-210.
The functions $S_i(t)$ describe source terms and the initial conditions are expressed as:
\begin{equation}
N_i(0)=N_{i,0}
\end{equation}
These equations can be treated as a decoupled system of equations in the form
\begin{equation}
\frac{N_i(t)}{dt}=-\frac{N_i(t)}{\tau_i}+\xi(t)
\end{equation}
that is readily solved as follows:
\begin{equation}\label{ft}
N_i(t) =N_i\ e^{-\frac{t}{\tau_i}} + \tau_i\langle \xi(t) \rangle_i
\end{equation}
where we used the notation
\begin{equation}
\langle \xi(t) \rangle_i= \frac{1}{\tau_i}\int_0^t e^{-\frac{t-t'}{\tau_i}}  \xi(t') dt'
\end{equation}
that can be thought of as a suitable average over an interval of time of the order of 
$\tau_i$ preceding the time  $t$. Some explicit expressions are given in the following.

The solution of the homogeneous system (i.e., with source terms set to zero)
with the boundary condition ${\overline N}_i(0)=N_{i,0}$ is:
\begin{equation}
\begin{array}{ll}
{\overline N}_1 (t)= & e^{-\frac{t}{\tau_1}} N_{1,0}  \\[1ex]
{\overline N}_2(t)= & e^{-\frac{t}{\tau_1}} N_{1,0}\,\frac{\epsilon_{21}}{1-\epsilon_{21}}
+e^{-\frac{t}{\tau_2}} \left(N_{2,0} -N_{1,0} \, \frac{\epsilon_{21}}{1-\epsilon_{21}}\right) \\[1ex]
{\overline N}_3(t)=  &e^{-\frac{t}{\tau_1}} N_{1,0}\, \frac{\epsilon_{31}}{(1-\epsilon_{21})(1-\epsilon_{31})}
-e^{-\frac{t}{\tau_2}}\left(N_{2,0} -N_{1,0}\, \frac{\epsilon_{21}}{1-\epsilon_{21}}\right) \frac{1}{1-\epsilon_{23}} +\\
&e^{-\frac{t}{\tau_3}}  \left[ N_{3,0} +  \left( N_{2,0}  -N_{1,0}\,\frac{\epsilon_{31}}{1-\epsilon_{31}}\right) 
\frac{1}{1-\epsilon_{23}}\right]\\
\end{array}
\end{equation}
where we used the notation
\begin{equation}
\epsilon_{ij}=\frac{\tau_i}{\tau_j}
\end{equation}
and considered the inequality $\tau_1\gg \tau_3 \gg \tau_2$.
The inhomogeneous (full) system admits the following solution 
satisfying the boundary condition $\delta N_i(0)=0$:
\begin{equation}
\begin{array}{l}
\delta N_1(t)=  \tau_1 \langle S_1(t) \rangle_1  \\[1ex]
\delta N_2(t)= \tau_2 \langle S_2(t)  +  \langle S_1(t) \rangle_1 \rangle_2  \\[1ex]
\delta N_3(t)=  \tau_3 \langle S_3(t)  +  \langle S_2(t)  + 
\langle S_1(t) \rangle_1 \rangle_2 \rangle_3  
\label{inhomogeneous}
\end{array}
\end{equation}
Thus, due to the linearity of the system, 
the most general  solution 
in this case is just 
\begin{equation}N_i(t)={\overline N}_i(t)+\delta N_i(t)\end{equation}

As an application, we discuss the case when we do not 
have sources of Bismuth or Polonium but only of Lead 
(this corresponds to the practical situation in which the Radon
content of the detector increases with time, since Radon-222 converts 
rapidly in Lead-210).
We consider the simplest case, namely when the source is a constant:
\begin{equation}
S_1(t)=S_1,\ S_2(t)=S_3(t)=0
\end{equation}
In this assumption, eqs.(\ref{inhomogeneous}) can be explicitly
 calculated obtaining:
\begin{equation}
\begin{array}{l}
\delta N_1(t)=  S_1 \tau_1 (1 -e^{-\frac{t}{\tau_1}}) \\[1ex]
\delta N_2(t)=  S_1 \tau_2 \left(1 -e^{-\frac{t}{\tau_1}} \frac{1}{1-\epsilon_{21}}
+e^{-\frac{t}{\tau_2}} \frac{\epsilon_{21}}{1-\epsilon_{21}} \right) \\[1ex]
\delta N_3(t)=  S_1 \tau_3 \left[ 1 -e^{-\frac{t}{\tau_1}} \frac{1}{(1-\epsilon_{21})(1-\epsilon_{31})}
- e^{-\frac{t}{\tau_2}} \frac{\epsilon_{21} \epsilon_{23}}{(1-\epsilon_{21})(1-\epsilon_{23})} +
 e^{-\frac{t}{\tau_3}} \frac{\epsilon_{31} }{(1-\epsilon_{23})(1-\epsilon_{31})}  \right]\\[1ex]
\end{array}
\end{equation}
Now consider the fact $\tau_2\ll \tau_3\ll \tau_1$ and the limit $\tau_2\ll t\ll \tau_1$, which applies to the typical conditions of data taking, $t$ being on the scale of one year. We get 
the approximated solutions:
\begin{equation}
\begin{array}{l}
N_{\rm Pb}(t)\approx N_{\rm Pb,0} +  t \,  (S_{\rm Pb} - \frac{N_{\rm Pb,0}}{\tau_{\rm Pb}} ) \\[1ex]
N_{\rm Bi}(t)\approx  \frac{\tau_{\rm Bi}}{\tau_{\rm Pb}}\ N_{\rm Pb}(t)  \\[1ex]
N_{\rm Po}(t)\approx  N_{\rm Po,0}  \; \exp\left(-\frac{t}{\tau_{\rm Po}} \right)  + \tau_{\rm Po}\, \left\langle \frac{ N_{\rm Bi}(t)}{\tau_{\rm Bi}} \right\rangle_{\rm Po}
\\[1ex]
\end{array}
\end{equation}
where we used the notations adopted in the paper to increase readability.
If the Lead-210 source is sufficiently intense, i.e. $S_{\rm Pb} > N_{\rm Pb, 0}/\tau_{\rm Pb}$, 
we expect a linear increase of the  Bi-210 beta activity in the detector, $n_{\rm Bi}(t) = N_{\rm Bi}(t)/\tau_{\rm Bi}$. 
When the Lead-210 source can be neglected, we obtain instead $n_{\rm Bi}(t)\simeq {\rm const}$.


\newpage

\section*{\sf  References}
\def\refname{

\end{document}
\vskip-1cm}
\baselineskip=1.15em


\begin{thebibliography}{99}
\bibitem{Kamland} 
KamLAND Collaboration, Phys. Rev. Lett. 90 (2003) 021802. \\
 KamLAND Collaboration, Phys. Rev. Lett. 100 (2008) 221803.

\bibitem{BXfirst} C. Arpesella et al., Borexino Collaboration, Phys. Lett. B 658 (2008) 101.

\bibitem{SNO+}
SNO+ Collaboration, Prog.~Part.~Nucl.~Phys. 57 (2006) 150.

\bibitem{LENA}
M.~Wurm {\it et al.},
  Acta Phys.\ Polon.\  B  41 (2010)  1749 
  [arXiv:1004.3474 [physics.ins-det]].

\bibitem{bethe}
C.F. Von Weizs\"acker, Physikalische Zeitschrift 39 (1938)  633.\\
A.H. Bethe, Physical Review 55 (1939) 434.


\bibitem{castellanietal}
R. Kippenhahn and A. Weigert, ``Stellar Structure and Evolution'', 
Springer-Verlag,  1991. \\
V. Castellani, ``Astrofisica Stellare'', Zanichelli, 1985.


\bibitem{noiS14}
S.~Degl'Innocenti, G.~Fiorentini, B.~Ricci and F.L.~Villante,
  Phys.\ Lett.\  B 590 (2004) 13.

\bibitem{Carlos}  A.M.~Serenelli,
  Astrophys.\ Space Sci.\  328 (2010) 13


\bibitem{Gonzalez} M.C.~Gonzalez-Garcia, M.~Maltoni and J.~Salvado, JHEP 1005 (2010) 072.

\bibitem{StrumiaVis}
  A.~Strumia and F.~Vissani,
  arXiv:hep-ph/0606054.

\bibitem{PRLBorex}
Borexino Collaboration,
  Phys.\ Rev.\ Lett.\  101 (2008) 091302.


\bibitem{as05}
M. Asplund, N. Grevesse and A.J. Sauval, Astronomical Society of the
Pacific Conference Series 336 (2005) 25,  T.G. Barnes and F.N. Bash
editors.

\bibitem{as09}
M. Asplund, N. Grevesse, A.J. Sauval and P. Scott, Ann. Rev. Astr. Astroph. 47 (2009) 481.

\bibitem{caffau09}
E.~Caffau, H.G.~Ludwig, M.~Steffen, B.~Freytag and P.~Bonifacio,
  arXiv:1003.1190 [astro-ph.SR].

\bibitem{gs98}
N. Grevesse and A.J. Sauval, Space Science Rev. 85 (1998) 161.

\bibitem{Basu}
 S.~Basu and H.M.~Antia,
  Phys.\ Rept.\  457 (2008) 217.

\bibitem{noi}
F.L.~Villante and B.~Ricci,
  Astrophys.\ J.\  714 (2010) 944.\\
F.L.~Villante,
  Astrophys.\ J.\  724 (2010) 98 .

\bibitem{TesteraPhysun2010} G. Testera, talks at PHYSUN 2010 available online at {\tt http://physun2010.mi.infn.it/}.

\bibitem{PDG} Review of Particle Physics, K. Nakamura et al., (Particle Data Group), J. Phys. G 37 (2010) 075021. 

\bibitem{Gatti} E.~Gatti, F.D.~Martini, ``A new linear method of discrimination between elementary particles in liquid scintillators'', IAEA Wien (1965) 265.

\end{thebibliography}
